\documentclass{pasa}%
\usepackage{graphicx}

\title{A broadband radio view of transient jet ejecta in the black hole candidate X-ray binary MAXI J1535--571}
\author[J. Chauhan et al.]{J. Chauhan$^{1}$\thanks{E-mail: j.chauhan@student.curtin.edu.au (JC)}, J.~C.~A. Miller-Jones$^{1}$\thanks{James.Miller-Jones@curtin.edu.au (JCAM-J)}, G.~E. Anderson$^{1}$, A. Paduano$^{1}$, M. Sokolowski$^{1}$, C. Flynn$^{2}$, P.~J. Hancock$^{1}$, N. Hurley-Walker$^{1}$, D.~L. Kaplan$^{3}$, T.~D. Russell$^{4,5}$, A. Bahramian$^{1}$, S.~W. Duchesne$^{1}$, D. Altamirano$^{6}$, S. Croft$^{7,8}$, H.~A. Krimm$^{9}$, G.~R. Sivakoff$^{10}$, R. Soria$^{11,12}$, C.~M. Trott$^{1}$, R.~B. Wayth$^{1}$, V. Gupta$^{2}$, M. Johnston-Hollitt$^{1,13}$, S.~J. Tingay$^{1}$
\\
\affil{$^{1}$International Centre for Radio Astronomy Research -- Curtin University, GPO Box U1987, Perth, WA 6845, Australia}%
\affil{$^{2}$Centre for Astrophysics and Supercomputing, Swinburne University of Technology, Mail H30, PO Box 218, VIC 3122, Australia}%
\affil{$^{3}$Department of Physics, University of Wisconsin–Milwaukee, Milwaukee, WI 53201, USA}%
\affil{$^{4}$INAF/IASF Palermo, via Ugo La Malfa 153, I-90146 Palermo, Italy}%
\affil{$^{5}$Anton Pannekoek Institute for Astronomy, University of Amsterdam, NL-1098 XH Amsterdam, the Netherlands}%
\affil{$^{6}$School of Physics and Astronomy, University of Southampton, B46, Southampton, SO17 1BJ, UK}%
\affil{$^{7}$Astronomy Department, University of California, Berkeley, 501 Campbell Hall 3411, Berkeley, CA 94720, USA}%
\affil{$^{8}$SETI Institute, 189 N Bernardo Ave \#200, Mountain View, CA 94043, USA}%
\affil{$^{9}$National Science Foundation, 2415 Eisenhower Avenue, Alexandria, VA 22314, USA}%
\affil{$^{10}$Department of Physics, University of Alberta, CCIS 4-181, Edmonton, AB T6G 2E1, Canada}%
\affil{$^{11}$College of Astronomy and Space Sciences, University of the Chinese Academy of Sciences, Beijing 100049, China}%
\affil{$^{12}$Sydney Institute for Astronomy, The University of Sydney, Sydney, NSW 2006, Australia}%
\affil{$^{13}$Curtin Institute for Computation, Curtin University, GPO Box U1987, Perth, 6845, WA, Australia}%
}%

\jid{PASA}
\doi{xxx/pas.\the\year.xxx}
\jyear{\the\year}

\usepackage{aas_macros}
\usepackage{hyperref} 
\hypersetup{colorlinks,citecolor=blue,linkcolor=blue,urlcolor=blue}
\usepackage[utf8]{inputenc}
\usepackage[explicit]{titlesec}
\titleformat{\paragraph}[display]{\normalfont\small\itshape}{\theparagraph. #1}{-1.5em}{}
\usepackage[T1]{fontenc}
\usepackage{ae,aecompl}
\usepackage{multicol}
\usepackage{longtable}
\usepackage{supertabular}
\usepackage{epsfig}
\usepackage{lineno}
\usepackage{lscape}
\usepackage{rotating}
\usepackage{natbib}
\usepackage{graphics}
\usepackage{gensymb}
\usepackage{graphicx}    
\usepackage{amsmath}    
\usepackage{amssymb}    
\usepackage{float}
\usepackage{xspace}
\usepackage[english]{babel}

\usepackage[dvipsnames]{xcolor}
\usepackage[nolist,nohyperlinks]{acronym}
\begin{document}

\begin{frontmatter}
\maketitle

\begin{abstract}
We present a broadband radio study of the transient jets ejected from the black hole candidate X-ray binary MAXI J1535--571, which underwent a prolonged outburst beginning on 2 September 2017. We monitored MAXI J1535--571 with the Murchison Widefield Array (MWA) at frequencies from 119 to 186 MHz over six epochs from 20 September to 14 October 2017. The source was quasi-simultaneously observed over the frequency range 0.84--19 GHz by UTMOST (the upgraded Molonglo Observatory Synthesis Telescope), the Australian Square Kilometre Array Pathfinder (ASKAP), the Australia Telescope Compact Array (ATCA), and the Australian Long Baseline Array (LBA). Using the LBA observations from 23 September 2017, we measured the source size to be $34\pm1$\,mas. During the brightest radio flare on 21 September 2017, the source was detected down to 119 MHz by the MWA, and the radio spectrum indicates a turnover between 250 and 500\,MHz, which is most likely due to synchrotron self-absorption (SSA). By fitting the radio spectrum with a SSA model and using the LBA size measurement, we determined various physical parameters of the jet knot (identified in ATCA data), including the jet opening angle ($\phi_{\rm op} = 4.5\pm1.2^{\circ}$) and the magnetic field strength ($B_{\rm s} = 104^{+80}_{-78}$ mG). Our fitted magnetic field strength agrees reasonably well with that inferred from the standard equipartition approach, suggesting the jet knot to be close to equipartition. Our study highlights the capabilities of the Australian suite of radio telescopes to jointly probe radio jets in black hole X-ray binaries via simultaneous observations over a broad frequency range, and with differing angular resolutions. This suite allows us to determine the physical properties of X-ray binary jets. Finally, our study emphasizes the potential contributions that can be made by the low-frequency part of the Square Kilometre Array (SKA-Low) in the study of black hole X-ray binaries.
\end{abstract}


\end{frontmatter}

\section{INTRODUCTION}
\label{sec:intro}
Stellar-mass black holes in X-ray binaries allow us to probe the fundamental processes of accretion and ejection \citep{Remillard2006, Fender2014}, because they evolve on humanly-observable timescales of months to years.  In contrast, their more massive analogues, the supermassive black holes, typically evolve much more slowly.

Black hole X-ray binaries (BH XRBs) spend most of their time in quiescence, where the X-ray luminosity is $<10^{-5.5}$ times the Eddington luminosity \citep{Plotkin2017}. In quiescence, the inner mass-accretion rate is lower than the mass transfer rate from the donor star. That builds up the matter in the accretion disk until the surface density is high enough to trigger a thermal-viscous instability, causing the disk to move into a hot, bright state observed as an outburst \citep[e.g.;][]{van_Paradijs1984, Dubus2001, Lasota2001, Lasota2008, Coriat2012}, the duration of which varies from a few weeks to a few years \citep[e.g.,][]{Tetarenko2016}.

During the outburst, along with the increased rate of mass transfer through the accretion disk, BH XRBs also show ejection of matter in the form of powerful jets \citep[e.g.,][]{Fender2006, Fender2014}, which extract a considerable amount of energy from the accretion flow and deposit it into their surroundings \citep[e.g.,][]{Gallo2005, Tetarenko2018}. Although it is believed that accretion and ejection are interconnected \citep[e.g.,][]{Tananbaum1972, Harmon1995, Hannikainen1998, Fender2004}, the jet launching and collimation mechanisms are still not fully understood. Owing to their rapid evolution, studies of BH XRBs can allow us to probe the causal connection between changes in the accretion flow and subsequent changes in the jets.

Throughout their outburst cycles, BH XRBs show characteristic X-ray spectral states, namely hard, hard-intermediate (HIMS), soft-intermediate (SIMS), and soft \citep{Remillard2006, Belloni2010}. These states are believed to be related to the geometry of the accretion flow, and a particular X-ray spectral state is connected to a specific kind of radio ejection \citep[e,g.,][]{Vadawale2003, Fender2004, Remillard2006, Belloni2010}. Two types of jets are observed, a compact jet, and a transient jet, which can be distinguished on the basis of their spectral indices ($\alpha$, where $S_{\nu}\propto\nu ^{\alpha}$; $S_{\nu}$ is the radio flux density and $\nu$ is the frequency) and morphology \citep[e.g.,][]{Fender2006}.

Optically thick, flat or slightly inverted spectra ($0\lesssim\alpha\lesssim0.6$) from partially self-absorbed, steady, compact jets \citep{Blandford1979} are observed during the hard X-ray spectral state \citep{Hjellming1988, Harmon1995, Fender2001}, which are associated with a dominant optically thin (hard power-law) component in the X-ray spectrum \citep{Thorne1975, Fender2004}. Due to their compact nature, these jets have been directly resolved in a few XRBs \citep[e.g.,][]{Stirling2001, Miller-Jones2021}.

When the X-ray luminosity increases to $\gtrsim 10^{37}$\,erg s$^{-1}$, the inner edge of the accretion disk is believed to move in towards the compact object \citep{Esin1997, Tang2011}. We observe transient jets during the intermediate X-ray spectral states, near the peak of the outburst. This usually occurs when a BH XRB is transitioning from the hard to the soft X-ray spectral state \citep{Vadawale2003, Fender2004}, as the X-ray spectrum becomes progressively more dominated by the thermal emission from the accretion disk \citep{Remillard2006}. However, in the case of Cygnus X--3, the opposite behaviour has been observed, where the strongest radio flares are observed when the source transitions from the soft to the hard X-ray spectral state. This difference is ascribed to the strong stellar wind of the Wolf-Rayet donor star, which during the soft state (when the jets are off) fills in the channel evacuated by the jets.  When the jets turn on again during the transition back to the hard state, they run into this dense medium, creating bright radio flares \citep[e.g.,][]{Koljonen2013, Koljonen2018}.

The transient jets are bright, relativistically-moving, discrete, expanding ejecta \citep[e.g.,][]{Mirabel1994, Hjellming1995, Tingay1995, Miller-Jones2012}. They are believed to be ejected on both sides of the compact object (relative to the accretion disc). The jets often appear one-sided, likely owing to Doppler boosting,  although the possibility of intrinsic asymmetry has been raised \citep[e.g.,][]{Fendt2013}. These ejecta are optically thin (above a certain frequency), having a steep radio spectrum ($-1\leqslant\alpha\leqslant-0.2$). A turnover in the radio spectrum is sometimes observed \citep[e.g.,][]{Miller-Jones2004, Chandra2017}, which moves to lower frequencies as the emitting region expands \citep{Laan1966}. Such a spectral turnover could either be caused by synchrotron self-absorption (SSA), or by free-free absorption (FFA) by thermal plasma \citep[see][and references therein]{Gregory1974, Miller-Jones2004}. In the case of SSA, the physical parameters of the jet set the turnover frequency, which scales with the magnetic field strength and the radius of the emitting region.

Australia has a suite of complementary radio telescopes that enable us to study both types of radio jets in BH XRBs over a broad frequency range (0.08--110 GHz).  This suite comprises the Murchison Widefield Array \citep[MWA:][]{Tingay2013, Wayth2018}, UTMOST \citep[the upgraded Molonglo Observatory Synthesis Telescope\footnote{As noted by \citet{Bailes2017}, UTMOST is not an acronym.};][]{Bailes2017}, the Australian Square Kilometre Array Pathfinder (ASKAP; \citealt{Hotan2014, Hotan2021}), the Australia Telescope Compact Array \citep[ATCA;][]{Frater1992, Wilson2011}, and the Australian Long Baseline Array \citep[LBA;][]{Preston1989, Preston1993, Jauncey1994}. This data set represents the first exploitation of the combined capabilities of all these telescopes for the purposes of studying jets from X-ray binaries.

MAXI J1535--571 underwent a prolonged outburst beginning on 2 September 2017 \citep{Negoro2017a, Kennea2017}, which was first detected by the Monitor of All-sky X-ray Image\footnote{\url{http://maxi.riken.jp/top/index.html}} \citep[{\it MAXI};][]{Matsuoka2009} and the Neil Gehrels Swift Observatory\footnote{\url{https://swift.gsfc.nasa.gov/}} \citep[{\it Swift}/BAT;][]{Gehrels2004, Barthelmy2005}. The outburst discovery was followed by a multi-wavelength monitoring campaign by the XRB community \citep[e.g.,][]{Dincer2017, Russell2017, Miller2018, Parikh2019,Russell2020}. MAXI J1535--571 underwent a bright radio flaring event reaching $\sim590$ mJy at 1.34 GHz around 21 September 2017 \citep{Chauhan2019b, Russell2019}.

Most of the physical parameters of MAXI J1535--571 are still uncertain. Recently, the H\,{\sc i} absorption line was used to determine a source distance of $4.1^{+0.6}_{-0.5}$\,kpc \citep{Chauhan2019b}, implying that at the peak of its outburst MAXI J1535--571 was accreting near to the Eddington limit. \citet{Russell2019} performed an extensive analysis of MAXI J1535--571 in the radio frequency band (5.5--19 GHz) and detected a discrete transient jet knot moving away from the compact object at a speed of $\geqslant0.69$c. The authors also constrained the jet inclination angle (at the time of ejection) to be $\leqslant45^{\circ}$. 

In this study, we report on our MWA, UTMOST and LBA observations of the transient radio jets from MAXI J1535--571 during September and October 2017, and combine these with previously-published ATCA \citep{Russell2019, Russell2020} and ASKAP \citep{Chauhan2019b} data to determine the physical properties of the jets. In Section~\ref{sec:observations}, we present detailed information on our observations and data reduction techniques, followed by the results in Section~\ref{sec:Results}. Section~\ref{sec:Radiospec} presents our radio spectral analysis and Section~\ref{sec:discussion} discusses the significance of our results. In Section~\ref{sec:conclusions} we provide the conclusions from our study.

\section{Observations and Data reduction}
\label{sec:observations}
\subsection{MWA}
During the 2017--2018 outburst of MAXI J1535--571, the MWA observed the source over six epochs from 20 September to 14 October 2017. In its Phase I, the MWA had an angular resolution of $\sim3$ arcmin at 154 MHz \citep{Tingay2013}. During the Phase II major upgrade in 2017, the angular resolution was increased by a factor of $\sim2$, and the sensitivity by a factor of $\sim4$, due to the associated reduction in the confusion noise \citep{Wayth2018}. At the time of our observations, the MWA was being upgraded from Phase I to Phase II, changing the resolution and sensitivity in each observation. The observations from the third, fourth and fifth epochs (26, 28 and 29 September 2017) have been excluded from this study due to the small number ($<64$) of available tiles, leading to poor data quality, and low resolution and sensitivity.

Our observations were carried out in three frequency bands centred at 119, 154 and 186 MHz, with 30.72\,MHz of bandwidth at each frequency. We observed MAXI J1535--571 for around 26 minutes in each frequency band, except on 14 October 2017 when we were restricted to 8 minutes per band (Table \ref{tab:tab1}). Our observations comprised 13 individual 2-minute snapshots, followed by a 112-s observation of a bright calibrator source, Hercules\,A. Further details of the MWA observations can be found in Table \ref{tab:tab1}.

We initially processed the raw visibility data with the {\tt COTTER} software \citep{Offringa2015}, which also excises the channels contaminated by radio-frequency inference (RFI) using the in-built {\tt AOFlagger} \citep{Offringa2012} tool, and converts the data to measurement set format. We then calibrated the data with the Common Astronomy Software Application \citep[{\tt CASA} v5.1.2-4:][]{McMullin2007}, using a bright, persistent extragalactic calibrator source (Hercules\,A). We made images using {\tt WSClean} \citep{Offringa2014}, employing Briggs weighting with a robust parameter of 0. We used the {\tt flux\_warp}\footnote{\url{https://gitlab.com/Sunmish/flux_warp}} software package \citep{Duchesne2020} to calibrate the flux density scale of each two-minute snapshot image using the persistent point sources from the GaLactic and Extragalactic All-sky MWA (GLEAM) catalogue \citep{Hurley-Walker2017} that could be identified in our image. The maximum correction found to be required to the absolute flux density scale was 10\%. Finally, we used the {\tt ROBBIE}\footnote{\url{https://github.com/PaulHancock/Robbie}} software package \citep{Hancock2019}, to correct the image for possible ionospheric distortions using the in-built {\tt fits\_warp}\footnote{\url{https://github.com/nhurleywalker/fits_warp}} software package \citep{Hurley-Walker2018}, and then create a mean image at each frequency band after stacking the individual two-minute snapshot images.

\subsection{UTMOST} 
The Molonglo Observatory Synthesis Telescope has been recently refurbished \citep{Bailes2017} via the UTMOST project. The telescope  now operates in a 31-MHz band centred at 835 MHz, although the sensitivity is not uniform across the band. The effective centre (weighted mean) of the band is 843 MHz. The telescope observes in a single circular polarization. It synthesises 351 narrow ($\approx 46$ arcsec) fanbeams with its East-West oriented arm, which tile out a wide (4\degree) field of view. Since June 2017 it has operated as a transit instrument, carrying out a Fast Radio Burst search and pulsar timing program.

Sources transit across the primary beam in 16 minutes on the equator, and in approximately 30 minutes at the declination of MAXI J1535--571, traversing the 351 fanbeams. The data processing backend writes these fanbeams as ``filterbank'' files to disk at 327 $\mu$sec resolution for 320 frequency channels, at a resolution of 98 kHz. In normal operations, these are decimated to 654 $\mu$sec and 40 $\times$ 0.7 MHz channels.

Transit observations of MAXI J1535--571 were obtained on 21, 26 and 27 September 2017, and all resulted in clear detections. For each observation, we measured the S/N of the source from the decimated filterbanks as it transited the fanbeam pattern, referencing it to sources of known flux density from the Molonglo Galactic Plane Survey \citep[undertaken at Molonglo prior to it becoming a transit instrument;][]{Murphy2007}.  These sources (MGPS 1541--5645, MGPS 1525--5709, MGPS 1533--5642 and MGPS 1532--5556, whose known flux densities are 282, 226, 544 and 1092 mJy respectively) are at very similar declinations, both ahead of and behind the source in right ascension. A modest fraction of the data was affected by mobile handset traffic in small sub-sections of the observing band, and this was flagged and removed. Subtraction of the background was required as the source is in the Galactic plane and there are a number of weak sources around the target. These could be identified readily as they traverse the field of view at a declination-dependent rate. Analysis of the flux calibration sources showed that systematics dominated the error budget, and were of the order of 30 to 50\,\% of the flux density.

\subsection{LBA}
We observed MAXI J1535--571 with the Australian Long Baseline Array (LBA) on 23--24 September 2017, from 22:26 to 05:30 UTC (MJD $58020.08\pm0.14$) under project code V456.  The array comprised seven stations (the phased-up ATCA, Ceduna, Hobart, Katherine, the Tidbinbilla 70-m dish DSS43, Warkworth and Yarragadee), although not all antennas were present at all times.  We observed at a central frequency of 8.441\,GHz, with the full 64\,MHz of bandwidth split into four 16-MHz IF pairs.  We used the bright extragalactic calibrator source PKS 0537-441 as a fringe finder and bandpass calibrator, and the closer source PMN J1515--5559 ($3.03^{\circ}$ from MAXI J1535--571) as a phase reference calibrator.  We used a 5-minute phase referencing cycle time, spending 3.5\,min on the target and 1.5\,min on the calibrator in each cycle.  The data were correlated using the DiFX software correlator \citep{Deller2007,Deller2011}, and reduced according to standard procedures within the Astronomical Image Processing System \citep[AIPS, version 31DEC17;][]{Greisen2003}.

Since we used the phased ATCA as one of our LBA stations, the observations also yielded a stand-alone ATCA data set.  We reduced these data within {\tt CASA} (version 5.6.2), using the standard calibrator PKS 1934--638 as a bandpass calibrator and to set the flux density scale. The array was in its compact H168 configuration\footnote{\url{https://www.narrabri.atnf.csiro.au/cgi-bin/obstools/baselines2.cgi?array=h168}}, with a maximum baseline of 192\,m between the inner five antennas, and the sixth antenna located 4.4\,km away.  We imaged the stand-alone ATCA data using the inner five antennas only.  MAXI J1535--571 was significantly detected, and its flux density derived by fitting a point source in the image plane using the {\tt IMFIT} task in {\tt CASA}.

\subsection{ASKAP}
ASKAP monitored the 2017--2018 outburst of MAXI J1535--571 over seven different epochs, from 21 September to 2 October 2017 (see Table \ref{tab:tab1}). A detailed description of the ASKAP observations and data reduction is presented in \citet{Chauhan2019b}. All the early science observations were carried out with an ASKAP sub-array of twelve dishes at a central frequency of 1.34 GHz with a processed bandwidth of 192 MHz. In this sub-array, ASKAP has an angular resolution of $\sim30$ arcsec. We processed our early science data with the standard ASKAP data analysis software,  {\tt ASKAPsoft}\footnote{\url{http://www.atnf.csiro.au/computing/software/askapsoft/sdp/docs/current/index.html}} \citep[pipeline version 0.24.1;][]{Guzman2019}. We estimated the flux densities (and $1\sigma$ uncertainties) of the source from the continuum images by using the {\tt IMFIT} task in {\tt CASA} v5.1.2-4.

\subsection{ATCA}
ATCA densely monitored the complete 2017--2018 outburst of MAXI J1535--571 \citep{Russell2019, Parikh2019}. The source was observed over thirty-seven epochs between 5 September 2017 and 11 May 2018. To complement our lower-frequency monitoring with ASKAP and MWA, we focus our analysis on those ATCA observations taken on 21, 23, 27 and 30 September 2017 (included in Table \ref{tab:tab1}). For these four ATCA epochs, data were recorded at 5.5, 9.0, 17.0 and 19.0 GHz, with 2\,GHz of bandwidth at each central frequency. To determine the overall behaviour of MAXI J1535--571 during the radio flaring event, we included all ATCA observations between 15 September and 25 October 2017 in our multi-frequency light curve presented in Section~\ref{sec:multi_light}. For a detailed description of the full ATCA monitoring and data analysis, see \citet{Russell2019}.

\begin{table*}
 \centering
 \caption{Details of the radio observations of MAXI J1535--571 used in this paper.} 
\begin{center}
\scalebox{1.0}{%
\begin{tabular}{ |l|c|c|c|c|c|c|c| }
\hline
\hline
Observation & Observation & MJD$^{\mathrm{a}}$ & Exposure & Telescope & Central & Flux density$^{\mathrm{b}}$ & References \\
Start date & Start time & & time &  & frequency & $S_{0}$ &  \\
(dd-mm-yyyy) & (hh:mm:ss) & & (min) &  & (GHz) & (mJy) &  \\
  & (UTC) &  &  &  &  &  & \\
\hline
20-09-2017 & 07:15:50 & 58016.31 & $\phantom{1}$28 & MWA & 0.119 & $<201$ &  This work\\
(MWA Epoch 1) & 07:17:50 & 58016.31 & $\phantom{1}$28 & MWA & 0.154 & $<102$ &  This work\\
 & 07:13:50 & 58016.31 & $\phantom{1}$28 & MWA & 0.186 & $<84$ &  This work\\
\hline
 21-09-2017 & 07:11:58 & 58017.31 & $\phantom{1}$26 & MWA & 0.119 & $\phantom{6}152\pm41$ & This work\\
(MWA Epoch 2) & 07:13:50 & 58017.31 & $\phantom{1}$26 & MWA & 0.154 & $\phantom{6}172\pm17$ &  This work\\
 & 07:15:50 & 58017.31 & $\phantom{1}$26 & MWA & 0.186 & $\phantom{6}194\pm16$ &  This work\\
 & 05:21:24 & 58017.31 & $\phantom{1}$30 & UTMOST & 0.840 & $\phantom{10}500\pm160$ & This work\\
 & 03:06:32 & 58017.17 & 122 & ASKAP & 1.34$\phantom{1}$ & $579.6\pm2.1$ & $[1]$\\
 & 09:01:30  & 58017.46 & 100  & ATCA & 5.5$\phantom{10}$ & $150.4\pm0.1$ & $[2]$\\
 & 09:01:30 & 58017.46 & 100 & ATCA & 9.0$\phantom{10}$ & $121.3\pm2.0$ &  $[2]$\\
 & 08:33:50  & 58017.46 & 100 & ATCA & 17.0$\phantom{100}$ & $\phantom{1}91.8\pm0.1$ & $[2]$\\
 & 08:33:50 & 58017.46 & 100 & ATCA & 19.0$\phantom{100}$ & $\phantom{1}85.8\pm0.1$ & $[2]$\\
\hline
22-09-2017 & 01:30:02 & 58018.12 & 172 & ASKAP & 1.34$\phantom{1}$ & $156.1\pm1.9$ & $[1]$\\
 & 10:35:41 & 58018.50 & 181 & ASKAP & 1.34$\phantom{1}$ & $306.1\pm1.3$ &  $[1]$\\
\hline
23-09-2017 & 13:23:39 & 58019.58 & $\phantom{1}$61 & ASKAP & 1.34$\phantom{1}$ & $478.2\pm2.4$ & $[1]$\\
 & 12:22:30 & 58019.52 & $\phantom{1}$20 & ATCA & 5.5$\phantom{10}$ & $377.2\pm1.2$ & $[2]$\\
 & 12:22:30 & 58019.52 & $\phantom{1}$20 & ATCA & 9.0$\phantom{10}$ & $324.2\pm0.3$ &  $[2]$\\
 & 11:54:50 & 58019.52 & $\phantom{1}$30 & ATCA & 17.0$\phantom{100}$ & $240.2\pm0.4$ & $[2]$\\
 & 11:54:50 & 58019.52 & $\phantom{1}$30 & ATCA & 19.0$\phantom{100}$ & $223.2\pm0.5$ & $[2]$\\
  & 22:26:36 & 58020.08 & 212 & ATCA & 8.44$\phantom{1}$ & $333\pm1$ & This work\\
  & 22:26:36 & 58020.08 & 212 & LBA & 8.44$\phantom{1}$ & $\lesssim100^{c}$ & This work\\
\hline
26-09-2017 & 09:40:00 & 58022.41 & $\phantom{1}$30 & UTMOST & 0.840 & $\phantom{1}200\pm100$ & This work\\
\hline
27-09-2017 & 09:40:00 & 58023.41 & $\phantom{1}$30 & UTMOST & 0.840 & $100\pm50$ & This work\\
 & 09:35:10 & 58023.42 & $\phantom{1}$30 & ATCA & 5.5$\phantom{10}$ & $127.5\pm0.3\phantom{1}$ & $[2]$\\
 & 09:35:10 & 58023.42 & $\phantom{1}$30 & ATCA & 9.0$\phantom{10}$ & $114.3\pm0.2\phantom{1}$ &  $[2]$\\
 & 09:07:30 & 58023.41 & $\phantom{1}$38 & ATCA & 17.0$\phantom{100}$ & $95.2\pm0.2$ & $[2]$\\
 & 09:07:30 & 58023.41 & $\phantom{1}$38 & ATCA & 19.0$\phantom{100}$ & $90.6\pm0.3$ & $[2]$\\
\hline
30-09-2017 & 03:29:59 & 58026.23 & 241 & ASKAP & 1.34$\phantom{1}$ & $39.8\pm0.8$ &  $[1]$\\
 & 06:45:20 & 58026.29 & $\phantom{1}$20 & ATCA & 5.5$\phantom{10}$ & $29.4\pm0.2$ &   $[2]$\\
 & 06:45:20 & 58026.29 & $\phantom{1}$20 & ATCA & 9.0$\phantom{10}$ & $26.8\pm0.1$ &   $[2]$\\
 & 06:17:40 & 58026.29 & $\phantom{1}$30 & ATCA & 17.0$\phantom{100}$ & $23.0\pm0.1$ &  $[2]$\\
 & 06:17:40 & 58026.29 & $\phantom{1}$30 & ATCA & 19.0$\phantom{100}$ & $23.5\pm0.1$ &  $[2]$\\
\hline
01-10-2017 & 03:35:00 & 58027.23 & 241 & ASKAP & 1.34$\phantom{1}$ & $26.3\pm0.8$ &  $[1]$\\
\hline
02-10-2017 & 03:39:59 & 58028.24 & 241 & ASKAP & 1.34$\phantom{1}$ & $21.4\pm0.7$ &  $[1]$\\
\hline
14-10-2017 & 05:29:26 & 58040.23 & $\phantom{20}$8 & MWA & 0.119 & $<117$ &  This work\\
(MWA Epoch 6) & 05:31:26 & 58040.23 & $\phantom{20}$8 & MWA & 0.154 & $<66$ &  This work\\
 & 05:33:26 & 58040.23 & $\phantom{20}$8 & MWA & 0.186 & $\phantom{8}82\pm17$ &  This work\\
\hline
\hline
\end{tabular}}\\
\end{center}
\begin{flushleft}
$^{\mathrm{a}}$ Mid point of the observations.\\
$^{\mathrm{b}}$ $1\sigma$ errors are presented, calculated by adding in quadrature the $1\sigma$ rms noise in the image and the $1\sigma$ error on the Gaussian fit to the source. For MWA, we also add in quadrature a 10\% uncertainty on the flux density scale.\\ 
$^{\mathrm{c}}$ Source is significantly detected on short baselines, but measured flux density falls off with baseline length.\\
$[1]$ \citet{Chauhan2019b}; $[2]$ \citet{Russell2019}. \\
Note: All upper limits are given at the $3\sigma$ level.\\
\end{flushleft}
\label{tab:tab1}
\end{table*}

\section{Results}
\label{sec:Results}
The 2017--2018 outburst of MAXI J1535--571 was detected across a broad radio frequency band by our set of complementary Australian telescopes. Our monitoring campaign allowed us to track the evolution of a transient jet knot \citep[denoted as S2 by][]{Russell2019}.

\subsection{Outburst evolution}
\begin{figure*}[t!]
    \includegraphics[width=2.1\columnwidth]{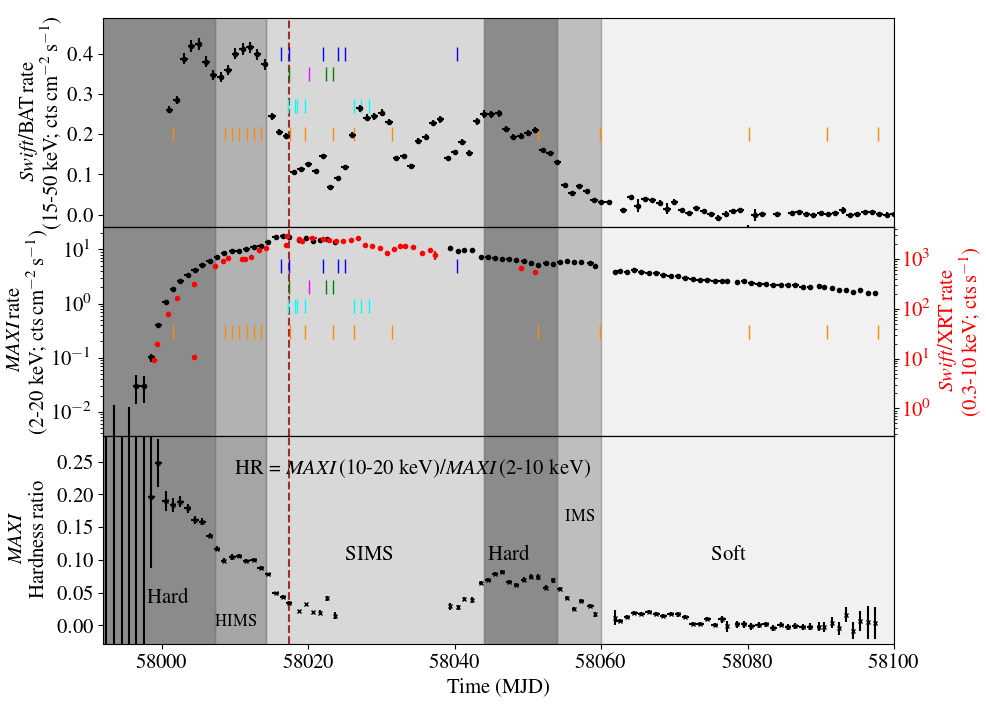}\vspace{-0.5em}
    \caption{ Top and  Middle panels: One-day averaged {\it Swift}/BAT, {\it Swift}/XRT and {\it MAXI} light curves of MAXI J1535--571 in the energy ranges 15.0--50.0 keV, 0.3--10.0 keV and 2.0--20.0 keV, respectively. Blue vertical lines highlight the dates of the MWA observations, green vertical lines indicate the UTMOST observations, and the LBA observation is denoted by the magenta vertical line. We also plot the dates of the ASKAP and ATCA observations with cyan and orange vertical lines, respectively \citep{Chauhan2019b, Russell2019}. The dashed brown vertical line indicates the 21 September 2017 observation when the source was brightest (reaching $\sim590$ mJy at 1.34 GHz in ASKAP observation), when we were able to measure a quasi--simultaneous broadband radio spectrum. Bottom panel: Variation of the hardness ratio (HR) calculated from {\it MAXI} on-demand public data. The HR is defined as the ratio of count rates in the 10.0--20.0 keV and 2.0--10.0 keV energy bands. Our observations were all taken during the soft--intermediate state.
    }
    \label{fig:fig1}
\end{figure*}

 In Fig.~\ref{fig:fig1}, we present the publicly-available one-day averaged X-ray light curves from {\it MAXI}, {\it Swift}/XRT, and {\it Swift}/BAT monitoring data of MAXI J1535--571, indicating the dates of the MWA, UTMOST, ASKAP, ATCA, and LBA observations. We have highlighted the X-ray spectral states from \citet{Tao2018} and \citet{Nakahira2018}, as described in \citet{Russell2019}. As is typical for BH XRBs \citep{Fender2004}, MAXI J1535--571 underwent a bright radio flaring event during its transition from the hard to the soft X-ray spectral state, peaking at $\sim590$ mJy at 1.34 GHz on 21 September 2017 \citep{Chauhan2019a, Russell2019}. Fig.~\ref{fig:fig1} shows that most of our radio monitoring data were taken over the peak of the outburst, when the source was in the soft--intermediate X-ray spectral state \citep{Chauhan2019b, Russell2019}.

\subsection{Low-frequency radio detections of MAXI J1535--571}
MAXI J1535--571 was detected at all three MWA frequencies (119, 154 and 186 MHz) on 21 September 2017 (our second MWA epoch), and this is the first transient BH XRB detected by MWA (to our knowledge). The source was not detected at any of the three frequencies on 20 September (the first MWA epoch), whereas on 14 October (the sixth MWA epoch), MAXI J1535--571 was detected only at 186 MHz, with $4.8\,\sigma$ significance (see Table \ref{tab:tab1}).

\begin{figure*}
    \includegraphics[width=1.0\columnwidth]{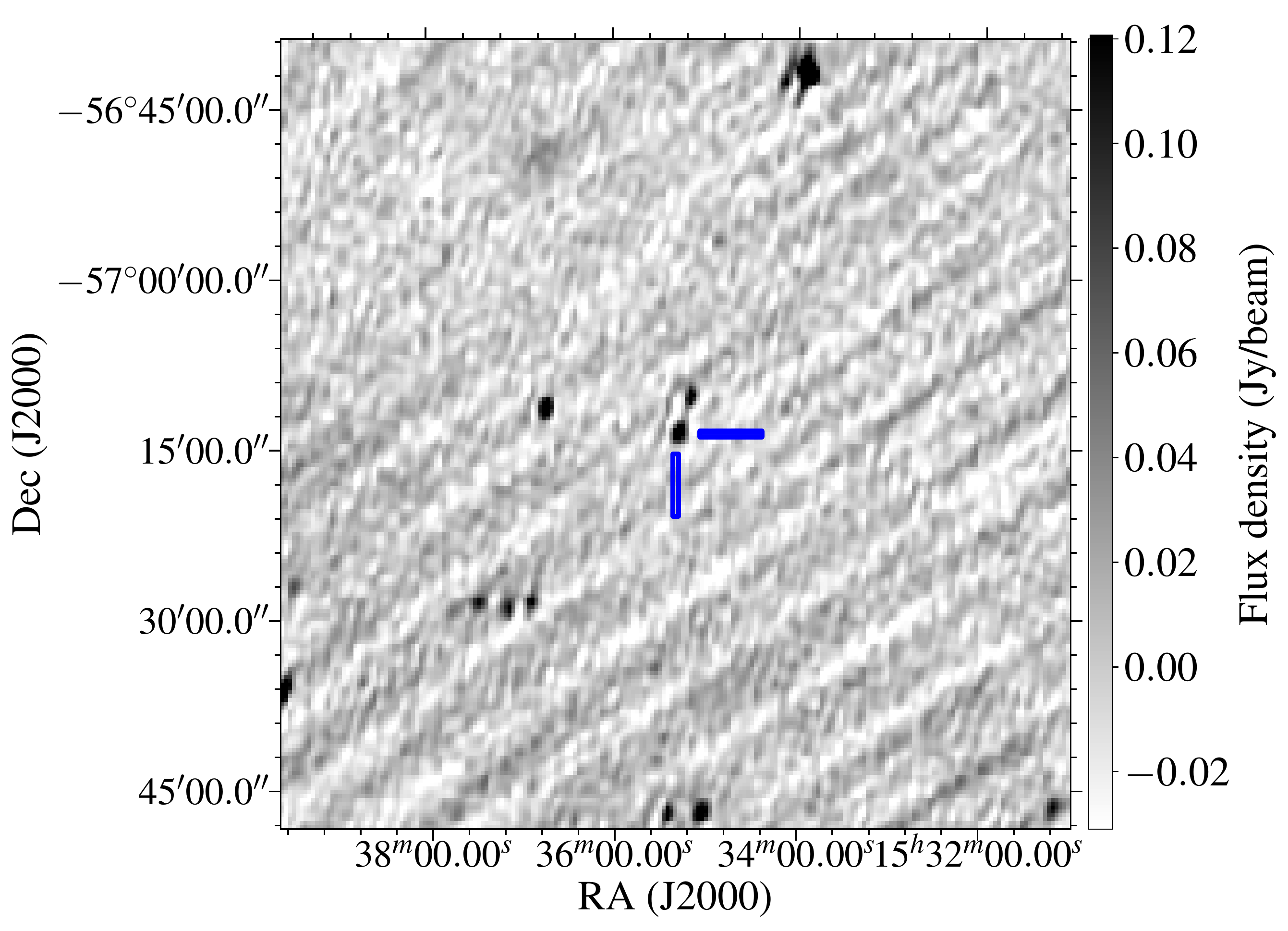}\vspace{0.5em}
    \includegraphics[width=1.0\columnwidth]{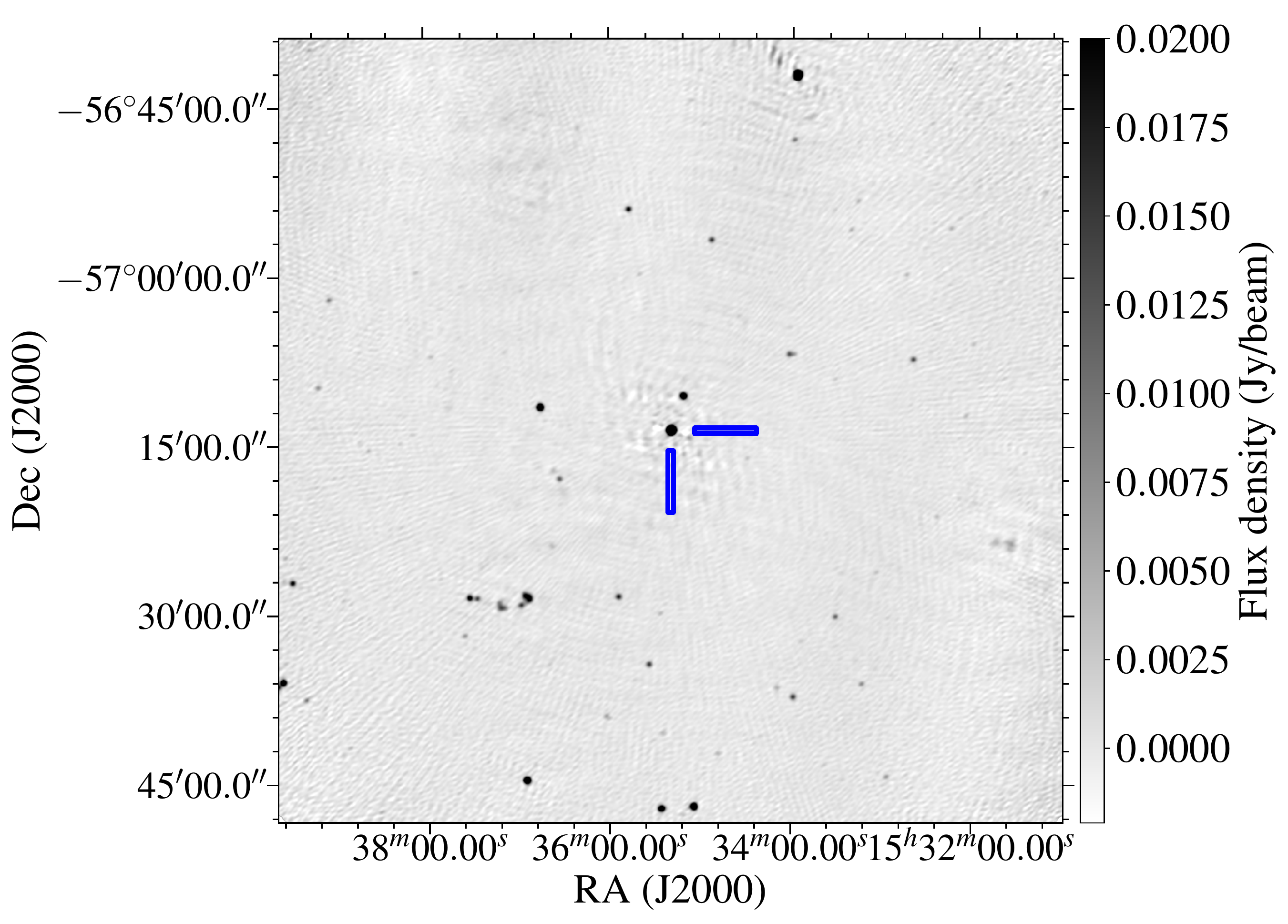}\vspace{0.5em}
    \caption{MWA (186 MHz; left panel) and ASKAP (1.34 GHz; right panel) continuum images of MAXI J1535--571 taken during the bright radio flare on 21 September 2017. The image is centred at the position of MAXI J1535--571 \citep[RA = 15:35:19.71, DEC = --57:13:47.58;][]{Russell2019} with a size of 1.16$\degree$ $\times$ 1.16$\degree$. In the MWA image, the diagonal stripes are sidelobes associated with PKS 1610-60 that is present to the south-east of MAXI J1535--571. These deconvolution artefacts are due to imperfect calibration resulting from the MWA's ongoing configuration change. MAXI J1535--571 is significantly detected in both images, and is indicated by the cross-hairs.
    }
    \label{fig:fig2}
\end{figure*}

In Fig.~\ref{fig:fig2}, we show the 186-MHz MWA continuum image of MAXI J1535--571 and the surrounding region for the 21 September 2017 observation, where the source was detected at $>10\,\sigma$ significance. For comparison, we also show the ASKAP 1.34-GHz continuum image of the same region on the same day, highlighting the difference in resolution of the two instruments.

MAXI J1535--571 was also detected in all three of the UTMOST observations, taken on 21, 26 and 27 September 2017.  We obtain 843-MHz flux density estimates of $500 \pm 160$, $200 \pm 100$ and $100\pm 50$ mJy (with the uncertainties dominated by systematics), respectively, indicating a clear fading of the source over the 6 day span of the observations (see Table \ref{tab:tab1}).

\subsection{Source size} \label{sec:size_LBA}
The stand-alone ATCA data observed as part of our LBA run on 23 September 2017 measured a flux density of $333\pm1$\,mJy at 8.44\,GHz (statistical errors only; to this should be added an additional systematic uncertainty on the flux density scale of 1--2\%). By contrast, the LBA data indicated a much lower level of emission, suggesting that a significant fraction of the ATCA emission was resolved out on the longer LBA baselines.  Clear fringes were only seen on the two shortest baselines (ATCA--Tidbinbilla, and Tidbinbilla--Hobart, respectively), with the flux density being $<100$\,mJy in both cases, and higher on the shorter baseline.  Since both baselines have almost the same orientation, this difference is a function only of baseline length, and hence places constraints on the source size scale.

The measured flux densities on these short baselines were seen to vary smoothly by up to a factor of 2 over the course of the observing run.  The simultaneous stand-alone ATCA data rule out this being due to intrinsic source variability, demonstrating that these baselines are probing the source structure. We used Difmap \citep{Shepherd1997} to project the visibilities along a range of different position angles, and found that when projected along a position angle of $125^{\circ}$ East of North \citep[the position angle of the moving jet knot S2 detected by][]{Russell2019}, they could be fit by a Gaussian of amplitude 333\,mJy (fixed to the measured ATCA flux density), with a width (standard deviation) of $6.1\pm0.1$\,M$\lambda$ (where M$\lambda=$ million wavelengths). Assuming that the jet knot brightness profile can be well approximated by a Gaussian, this corresponds to a size scale of $34\pm1$\,mas.  Our $uv$-coverage from these two baselines alone does not permit us to constrain the size scale in the perpendicular direction, so we assume that the knot can be modelled as a circular Gaussian of width $\theta _{\rm s} = 34\pm1$\,mas (corresponding to a physical size of $139^{+21}_{-17}$\,AU, calculated using the source distance of $4.1^{+0.6}_{-0.5}$\,kpc from \citealt{Chauhan2019b}).

This size constraint can be used to determine parameters including the synchrotron minimum energy (Section \ref{sec:Radiospec_SSA}), the magnetic field strength (Section \ref{sec:magnetic_field}), and the opening angle of the jet (Section \ref{sec:opening_angle}), improving our understanding of the energetics of the transient jet ejection.

\subsection{Multi-frequency radio light curve} \label{sec:multi_light}
\begin{figure*}[t!]
    \includegraphics[width=2.15\columnwidth]{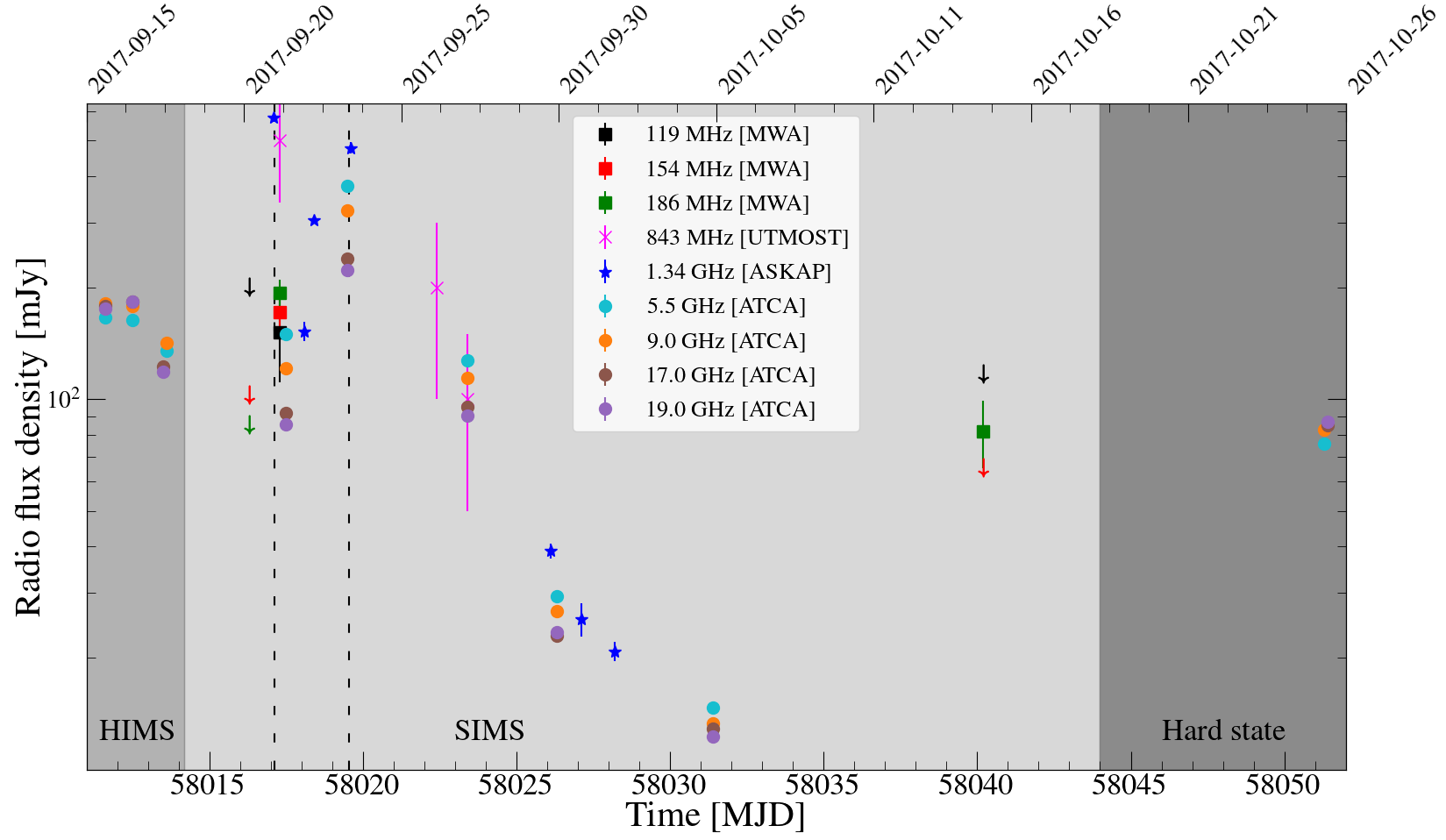}\vspace{-0.5em}
    \caption{Multi-frequency radio light curve of MAXI J1535--571. Solid squares, crosses, stars and circles correspond to MWA, UTMOST, ASKAP and ATCA observations, respectively. Different colours indicate different observing frequencies, as indicated by the plot legend. In the case of MWA non-detections, downward-pointing arrows represent $3\sigma$ upper limits on the radio flux density. The medium-dark shaded region highlights the HIMS, the SIMS is represented by the light shaded region, and the dark shaded region highlights the hard X-ray spectral state. At the start and end of the light curve, ATCA points indicate the quenching and reappearance of the compact jets \citep{Russell2019, Russell2020}. The two peaks are highlighted with the vertical dashed lines. The best-sampled date was 21 September (MJD 58017), during the first peak in the light curve.}
    \label{fig:fig3}
\end{figure*}

MAXI J1535--571 was observed by the Australian suite of radio telescopes during its radio flaring event in September 2017. The 0.12 -- 19 GHz radio light curve spanning from 15 September (MJD 58011) to 26 October (MJD 58052) is shown in Fig.~\ref{fig:fig3}. \citet{Russell2019, Russell2020} observed the compact jets beginning to quench around 17 September (MJD 58013.6), at the end of the HIMS, and just before the radio flaring event. In the light curve, we observed two clear peaks on 21 and 23 September 2017, in each of which the 1.34 GHz radio flux density exceeded $450$ mJy. The two peaks could arise from two separate ejection events. However, with no direct evidence for a second component from imaging studies (our LBA data or the ATCA data of \citealt{Russell2019}), it is also possible that the second peak in the light curve is due to re-brightening of the original synchrotron-emitting jet knot as it interacts with the surrounding medium. After the second peak, the radio flux density of the source gradually decayed at all frequencies, reaching $\sim13$ mJy in the 5.5--19.0 GHz frequency band on 5 October 2017 \citep{Russell2019}.

The MWA detection on 21 September 2017 (MJD 58017.31) coincides with the first radio flaring event observed from MAXI J1535--571 (see Fig.~\ref{fig:fig3}), in which the maximum flux density reached $580\pm2$ mJy at 1.34 GHz. However, the interpretation of the 186-MHz MWA detection on 14 October 2017 (MJD 58040.23) is less clear.  It either corresponds to the fading tail of the bright ejecta, or to low-frequency emission from the re-formed compact jets. Both interpretations are plausible. The transient ejecta would have a steep, optically thin spectrum, making them brightest at low radio frequencies. However, the MWA detection occurred in the SIMS (see Fig.~\ref{fig:fig3}), and the compact jets should already have reformed at GHz frequencies by the time of the subsequent HIMS, as seen in MAXI J1836--194 \citep{Russell2014}.  Furthermore, the MWA detection was at a similar flux density to the 5--19\,GHz ATCA detection of the re-formed compact jets on 25 October 2017 \citep{Russell2019}.

\subsection{Radio spectrum} \label{sec:R_spec}
\begin{figure}
    \includegraphics[width=\columnwidth]{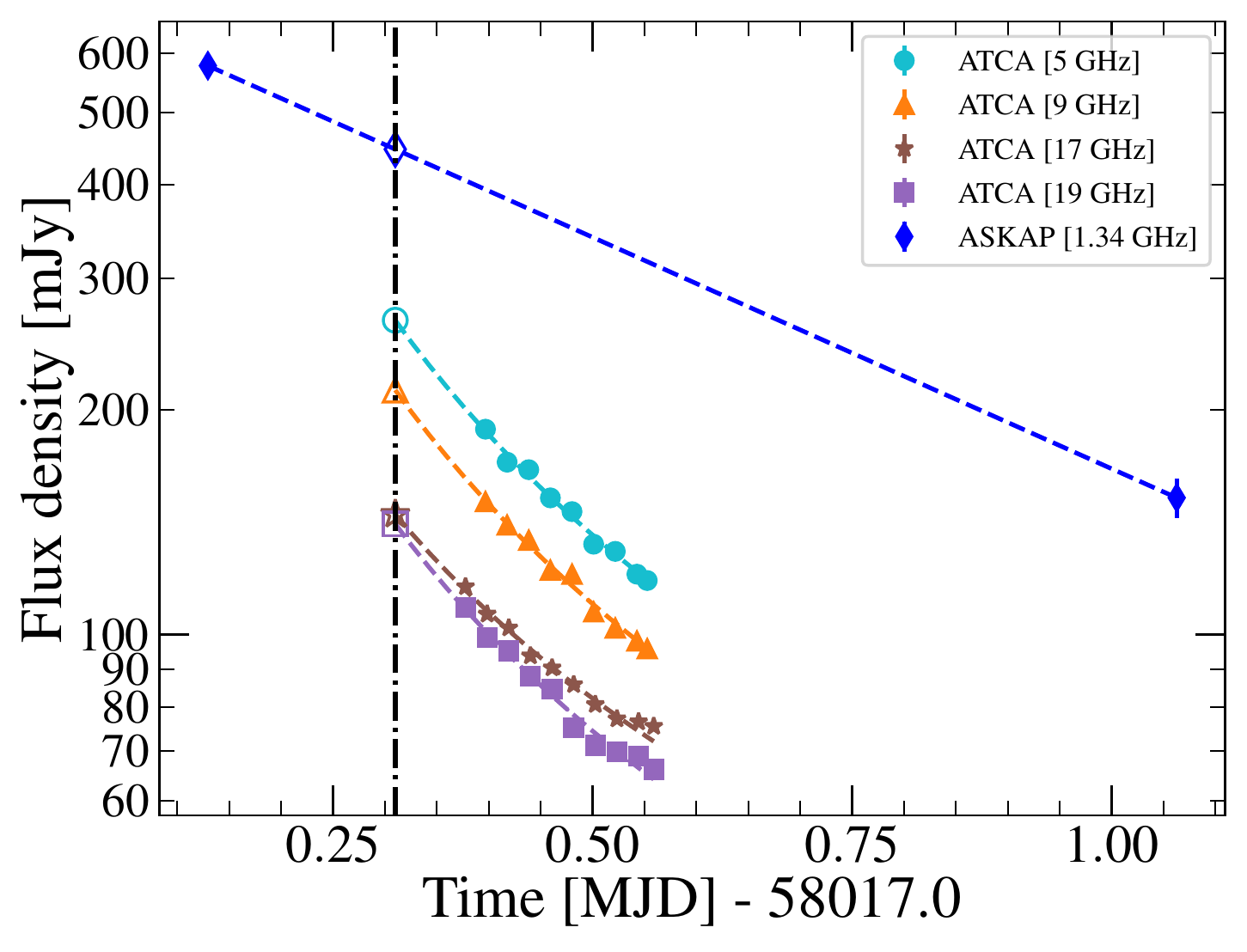}\vspace{0.5em}
    \caption{Short time-scale light curves of the ATCA observations on 21 September 2017. The vertical dash-dotted line indicates the time of the MWA observation. ASKAP/ATCA flux densities are interpolated/extrapolated to the time of the MWA observations, and shown with hollow markers. Each plotted symbol and its colour represents a different observing frequency, as indicated in the legend. The dashed lines represent the fitted power law models for the respective light curves (as described in Section~\ref{sec:R_spec}). By extrapolating/interpolating the flux density decays seen with ATCA and ASKAP, we reconstructed a strictly simultaneous radio spectrum at the time of the MWA observation.}
    \label{fig:fig4}
\end{figure}

\begin{figure}[t!]
    \includegraphics[width=\columnwidth]{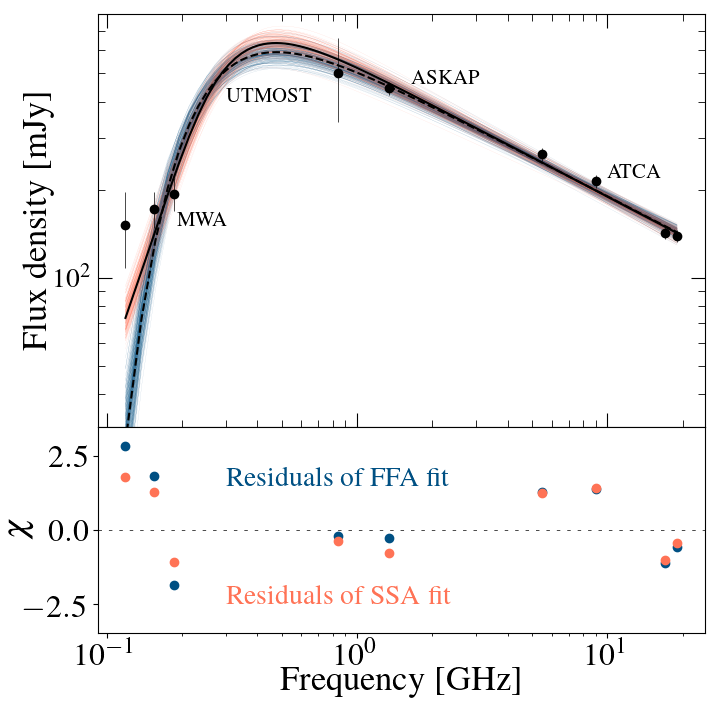}
    \caption{Top panel: Broadband radio spectrum of MAXI J1535--571 on 21 September 2017. The flux densities are from MWA (this work), UTMOST (this work), ASKAP \citep{Chauhan2019b} and ATCA \citep{Russell2019}, as indicated. The black solid line highlights the median of the posterior distribution for the SSA model (discussed in Section~\ref{sec:Radiospec_SSA}), whereas the black dashed line shows the median of the posterior distribution for the FFA model (described in Section~\ref{sec:Radiospec_FFA}). We have added systematic uncertainties on the flux densities measured by MWA (10\,\%), ASKAP (5\,\%) and ATCA \citep[a conservative 5\,\%, as appropriate for the higher frequencies;][]{Partridge2016}, to incorporate the cross-telescope uncertainties. The orange and blue traces show random draws from the posterior distributions of the best fits for the SSA and FFA models, respectively. Bottom panel: Residuals relative to the median of the posterior distributions for both the SSA and FFA models. The low-frequency residuals are lower for the SSA model. The low-frequency turnover allows us to estimate several of the physical parameters of the jet.}
    \label{fig:fig5}
\end{figure}

On 21 September 2017, the MWA spectrum was rising with frequency, whereas above 1 GHz, it was falling with frequency.  This implies a spectral turnover. However, the observations from MWA, ASKAP and ATCA were not strictly simultaneous. Given the rapid flux density variations during the flaring events, this non-simultaneity could bias our broadband radio spectrum.  We therefore broke the ASKAP and ATCA data into short time chunks of $\sim 20$ minutes each and fit them with a power law (Fig.~\ref{fig:fig4}), which we extrapolated back to the time of the MWA observations to reconstruct a simultaneous broadband radio spectrum. We also tried to fit the light curves with an exponential decay, but the $\chi^2$ values of the fits were lower for the power law fits, particularly at the higher frequencies of 17 and 19\,GHz. Our reconstructed simultaneous 0.12--19.0 GHz spectrum (Fig.~\ref{fig:fig5}) shows a clear turnover between 250 and 500 MHz, with a low-frequency spectral index of $\alpha _{l} = 0.91\pm0.60$ between 119 and 186 MHz, and a high-frequency spectral index of $\alpha _{h} = -0.44\pm0.01$ above 1 GHz.

\section{Radio spectral analysis} \label{sec:Radiospec}
To understand the physical scenario behind the observed low-frequency spectral turnover in our 21 September observation, we considered FFA by thermal plasma and SSA, both of which can produce a low-frequency turnover in the radio spectrum \citep[e.g.,][]{Gregory1974, Miller-Jones2004}. 

\subsection{Free-Free Absorption} \label{sec:Radiospec_FFA}
In the case of FFA, free electrons either in an external screen, or from thermal material mixed with synchrotron-emitting plasma, absorb the synchrotron photons in the presence of massive ions \citep{Kellermann1966}. For an ionised hydrogen cloud of length $l$ \citep[in parsec;][]{Mezger1967}, temperature $T_{\rm eq}$ ($\times\,10^{4}$ K), and electron number density $n_{\rm e}$ (in cm$^{-3}$), the optical depth ($\tau _{\nu}$) to FFA at frequency $\nu _{\rm GHz}$ (in GHz) can be expressed as \citep{Tingay2003}
\begin{equation}
\label{eq:optical_depth}
  \tau _\nu \approx 3.2 \times 10^{-7}\,T_{\rm eq}^{-1.35}\,\nu^{-2.1}_{\rm GHz} \int n_{\rm e}^{2}\,dl.
\end{equation}

To investigate the possibility of free-free absorption in our source, we considered the simplest scenario of a single homogeneous external absorbing screen of free electrons and scattering ions.  This predicts a flux density and optical depth that scale with frequency as
\begin{equation}
  S_\nu = S_{0}\,\nu^\alpha\,e^{-\tau _\nu},
\end{equation}
\begin{equation}
   \tau _\nu = \left(\frac{\nu}{\nu _{\rm p}}\right)^{-2.1},
\end{equation}
where $\nu _{p}$ is the frequency at which the optical depth becomes unity, $S_{0}$ is the flux density of the source at frequency $\nu _{p}$, and $\alpha$ is the spectral index of the synchrotron spectrum  \citep{Callingham2015}.

We tried to fit our observed broad-band radio spectrum (Fig.~\ref{fig:fig5}) with this FFA model (highlighted with the dashed line, blue traces, and blue residuals in Fig.~\ref{fig:fig5}), using a Bayesian approach, which provided best-fit estimates and parameter uncertainties. We created a Markov Chain Monte Carlo (MCMC) simulation, incorporated uniform priors of $\alpha$ = $-10$ -- 0, $S_{0}$ = 1 -- 1000\,mJy, and $\nu _{\rm p}$ = 0.05 -- 10\,GHz, and used the PyMC3\footnote{\url{https://docs.pymc.io/}} package developed by \citet{Salvatier2016}. The model estimated values and $1\sigma$ uncertainties of $\alpha = -0.44\pm0.02$, $S_{0} = 526\pm28$ mJy and $\nu _{\rm p} = 0.23\pm0.01$ GHz. The slope predicted by the model in the low-frequency regime is $2.96\pm0.16$, which is inconsistent with the measured slope of $\alpha _{l} = 0.91\pm0.60$ (Section~\ref{sec:R_spec}) at a high significance ($\lesssim3\sigma$). We therefore do not favour FFA as an explanation for the low-frequency turnover.

We also explored whether FFA from an external ionized gas region can be physically supported as a reasonable interpretation for the low-frequency turn over observed in the radio spectrum of MAXI J1535--571. We consider a hypothetical H\,{\sc ii} region along the line of sight to MAXI J1535--571, and calculate its expected H$_{\alpha}$ emission using the observed ranges of $n_{e}$ and $l$ for classical H\,{\sc ii} regions. Following the prescription given by \citet{Osterbrock2006}, we predict the total H$_{\alpha}$ luminosity of a hypothetical H\,{\sc ii} region to be in the range $\approx\,8\,\times\,10^{36}$ to $4\,\times\,10^{39}$\,erg\,s$^{-1}$. The H$_{\alpha}$ flux for a hypothetical H\,{\sc ii} region located close to the source distance of $\sim4.1$\,kpc (\citealt{Chauhan2019b}) is $\approx4$\,$\times$\,10$^{-9}$ to 2\,$\times$\,10$^{-6}$\,erg\,cm$^{-2}$\,s$^{-1}$, and the corresponding surface brightness is $\approx5$\,$\times$\,10$^{-15}$ to 3\,$\times$\,10$^{-11}$\,erg\,cm$^{-2}$\,s$^{-1}$\,arcsec$^{-2}$.

We analysed the data from the Southern H--Alpha Sky Survey Atlas\footnote{\url{http://amundsen.swarthmore.edu/SHASSA}} (SHASSA, \citealt{Gaustad2001, Finkbeiner2003}) to search for such H$\alpha$ emission. The $3\sigma$ upper limit on the mean surface brightness for a circular region of radius 0.35$^{\circ}$ centered on MAXI J1535--571 is $\sim2\,\times\,10^{-15}$\,erg\,cm$^{-2}$\,s$^{-1}$\,arcsec$^{-2}$, ruling out the presence of an H\,{\sc ii} region with the characteristics derived above.

We also calculated the predicted radio flux density at 5 GHz of the above hypothesized H\,{\sc ii} region using the prescription given by \citet{Caplan1986}, which is estimated to be 4 to 2000\,Jy. The corresponding radio surface brightness is $\approx5\,\times$\,10$^{-6}$ to 3\,$\times$\,10$^{-2}$\,Jy\,arcsec$^{-2}$. From the ATCA observation of MAXI J1535--571 on 22 February 2018, \citet{Russell2019} measured a deep $3\sigma$ upper limit on the 5.5-GHz radio flux density of 0.1\,mJy. The corresponding $3\sigma$ upper limit on the radio surface brightness is $3.4\times10^{-7}$\,Jy\,arcsec$^{-2}$, well below the estimated surface brightness for the hypothesized H\,{\sc ii} region. Both radio and H$\alpha$ observational constraints therefore argue against the presence of an H\,{\sc ii} region along the line of sight towards MAXI J1535--571, making it unlikely that FFA from an external screen is the main cause of the observed low-frequency turnover.

\subsection{Synchrotron Self-Absorption} \label{sec:Radiospec_SSA}
SSA is often suggested to be responsible for the low-frequency turnover in the radio spectrum of X-ray binaries \citep{Gregory1974, Seaquist1976}. SSA is an internal property of the source, and the turnover arises because below a certain frequency the electrons become optically thick to their own synchrotron radiation. In the case of SSA, at frequencies below the turnover ($< \nu _{p}$), where $\tau _{\nu} \gg 1$ (in the optically thick region), the synchrotron self-absorbed spectrum varies as $S_{\nu} \propto \nu^{5/2}$ \citep[e.g.,][]{Rybicki1979}. At frequencies above the turnover ($> \nu _{p}$), in the optically thin region ($\tau _{\nu} \ll 1$), the spectrum scales as $S_{\nu} \propto \nu^{\alpha}$ ($\alpha < 0$) \citep[e.g.][]{Laan1966, Rybicki1979}. Finally, the structure of the source defines the width of the turnover region. A synchrotron self-absorbed spectrum can be parametrised as
\begin{equation}
\label{eq:ssa}
  S_\nu = S_{0} \left(\frac{\nu}{\nu _{\rm p}}\right)^{-(\beta-1)/2} \left[\frac{1 - e^{-\tau _{\nu}^{'}}}{\tau _{\nu}^{'}}\right],
  \end{equation}
\begin{equation}
  \tau _{\nu}^{'}= \left(\frac{\nu}{\nu _{\rm p}}\right)^{-(\beta+4)/2},
\end{equation}
where $\beta$ is the power-law index of the electron energy distribution, and $\nu _{\rm p}$ represents the frequency where the source becomes optically thick \citep{Tingay2003, Callingham2015}. At $\nu _{\rm p}$, the mean free path of the synchrotron photons that scatter off the non-thermal electrons becomes comparable to the geometrical size of the synchrotron source (the jet knot). 

The SSA model provides a better fit in the low-frequency band as compared to the FFA model, as highlighted by the black solid line, orange traces and orange residuals in Fig.~\ref{fig:fig5}. After fitting the spectrum of MAXI J1535--571 with the SSA model in equation (\ref{eq:ssa}) and using uniform priors $\beta$ = 0 -- 10, $S_{0}$ = 1 -- 1000\,mJy, and $\nu _{p}$ = 0.05 -- 10\,GHz, we estimated (with $1\sigma$ uncertainties) $S_{0} = 882\pm56$ mJy, $\beta = 1.90\pm0.04$ and $\nu _{p} = 0.32\pm0.01$ GHz.

Using the aforementioned values together with our direct LBA size measurement, we can derive the minimum energy parameters without having to rely on the rise time of the radio flare to constrain the source size, or minimising the energy with respect to the source expansion rate, as recently proposed by \citet{Fender2019}.
We follow \citet{Fender2006}, who give an expression for the synchrotron minimum energy as
\begin{multline}
\label{eq:Emin}
    E_{\rm min} \sim 3\times10^{8}\,\eta^{4/7}\,\left(\frac{fV}{\rm cm^{3}}\right)^{3/7}\,\left(\frac{\nu _{\rm p}}{\rm Hz}\right)^{2/7}\\ \left(\frac{L_{\nu _{\rm p}}}{\rm erg\,s^{-1}\,Hz^{-1}}\right)^{4/7},
\end{multline}
where $V$ is the volume of the synchrotron emitting plasma, $f$ is the filling factor of the jet knot (assumed to be 1), $\nu _{\rm p}$ is the turnover frequency, and $L_{\nu _{\rm p}}$ is the monochromatic luminosity of the jet knot at the turnover frequency. $\eta = (1 + \beta_{\rm pe})$, where $\beta_{\rm pe}$ is the ratio of energy in protons to that in electrons, which we assume to be 0, such that $\eta=1$ \citep{Fender2006}.

The LBA observed MAXI J1535--571 on 23--24 September 2017 (MJD $58020.082\pm0.147$), $\sim3$ days after the peak of the radio outburst. Assuming a constant expansion speed and an ejection date of MJD $58010.8^{+2.7}_{-2.5}$ \citep{Russell2019}, we estimate the source size on 21 September to be $23.8^{+3.0}_{-2.8}$ mas. To calculate $V$, we assume the jet knot to be spherical, with a radius equal to that estimated source size at the known source distance \citep{Chauhan2019b}. From Equation \ref{eq:Emin}, we find a minimum energy value of $E_{\rm min}= 6.5\pm2.5\times10^{41}$ erg, which is at the high end of the range $10^{38} - 10^{42}$ erg reported from other XRBs such as V404 Cygni, Cygnus X--3, and GRS 1915+105 \citep[e.g.][]{Chandra2017, Fender2019}.

We also follow \citet{Fender2006} to calculate the minimum energy magnetic field strength, expressed as
\begin{multline}
\label{eq:Bmin}
    B_{\rm eq} \sim 1.6\times10^{4}\,\eta^{2/7}\,\left(\frac{fV}{\rm cm^{3}}\right)^{-2/7}\,\left(\frac{\nu _{\rm p}}{\rm Hz}\right)^{1/7}\\ \left(\frac{L_{\nu _{\rm p}}}{\rm erg\,s^{-1}\,Hz^{-1}}\right)^{2/7},
\end{multline}
The minimum--energy magnetic field strength is found to be $B_{\rm eq} = 40\pm5$ mG, which is in line with the limits (10--500\,mG) defined by \citet{Russell2019}, and comparable to canonical values for XRBs \citep{Fender2019}.

\section{DISCUSSION}  \label{sec:discussion}
Our study demonstrates the capabilities of the Australian suite of radio telescopes. As outlined in section \ref{sec:R_spec}, we detected a low-frequency turnover in the broad-band radio spectrum of MAXI J1535--571. While these are believed to be a common feature of transient XRB jets, limited high-cadence monitoring at low radio frequencies has meant that low-frequency turnovers have previously been detected in just five \citep{Seaquist1980, Seaquist1982, Miller-Jones2004, Chandra2017, Fender2019, Chauhan2019a} of the $\sim$60 known black hole candidate XRBs \citep{Corral-Santana2016, Tetarenko2016}. In Section~\ref{sec:Radiospec}, we found that the low-frequency turnover that we observed is most likely due to SSA. In the following subsections, we discuss the implications of our derived SSA model parameters.

\subsection{Magnetic field strength} \label{sec:magnetic_field}
Under the assumption that the low-frequency turnover is due to synchrotron self absorption, we can use our LBA measurement of the size of the jet knot to constrain the magnetic field strength $B_{\rm s}$ of the knot, as has often been done for extragalactic jets. This can be determined \citep{Marscher1983} as
\begin{equation}
\label{eq:B}
  B_s = 10^{-5}\,b(\alpha)\,\theta_{\rm s}^{4}\,\nu _{\rm p}^{5}\,S_{\rm 0}^{-2} \left[\frac{\delta_{\rm bp}}{1+z}\right]\quad{\rm G},
\end{equation}
where $\theta_{s}$ is the angular size of the synchrotron emitting region in mas, $S_{0}$ is the radio flux density in Jy at the self-absorption turnover frequency $\nu _{\rm p}$ (measured in GHz), and $\delta_{\rm bp} = [\Gamma(1-\beta\cos{i})]^{-1}$ is the Doppler factor of the jet, with {\it i} being the inclination angle of the jet axis to the line of sight, $\beta\left(=\frac{v}{c}\right)$ the jet speed, and $\Gamma=[1-{\beta^{2}}]^{-1/2}$ the bulk Lorentz factor of the jet. The quantity $b(\alpha)$ is a slowly varying function of the high-frequency spectral index $\alpha$, which has a value of $\approx3.4$ for $\alpha$ = --0.6. For a Galactic object the redshift $z$ can be set to 0.

\begin{figure}
    \includegraphics[width=\columnwidth]{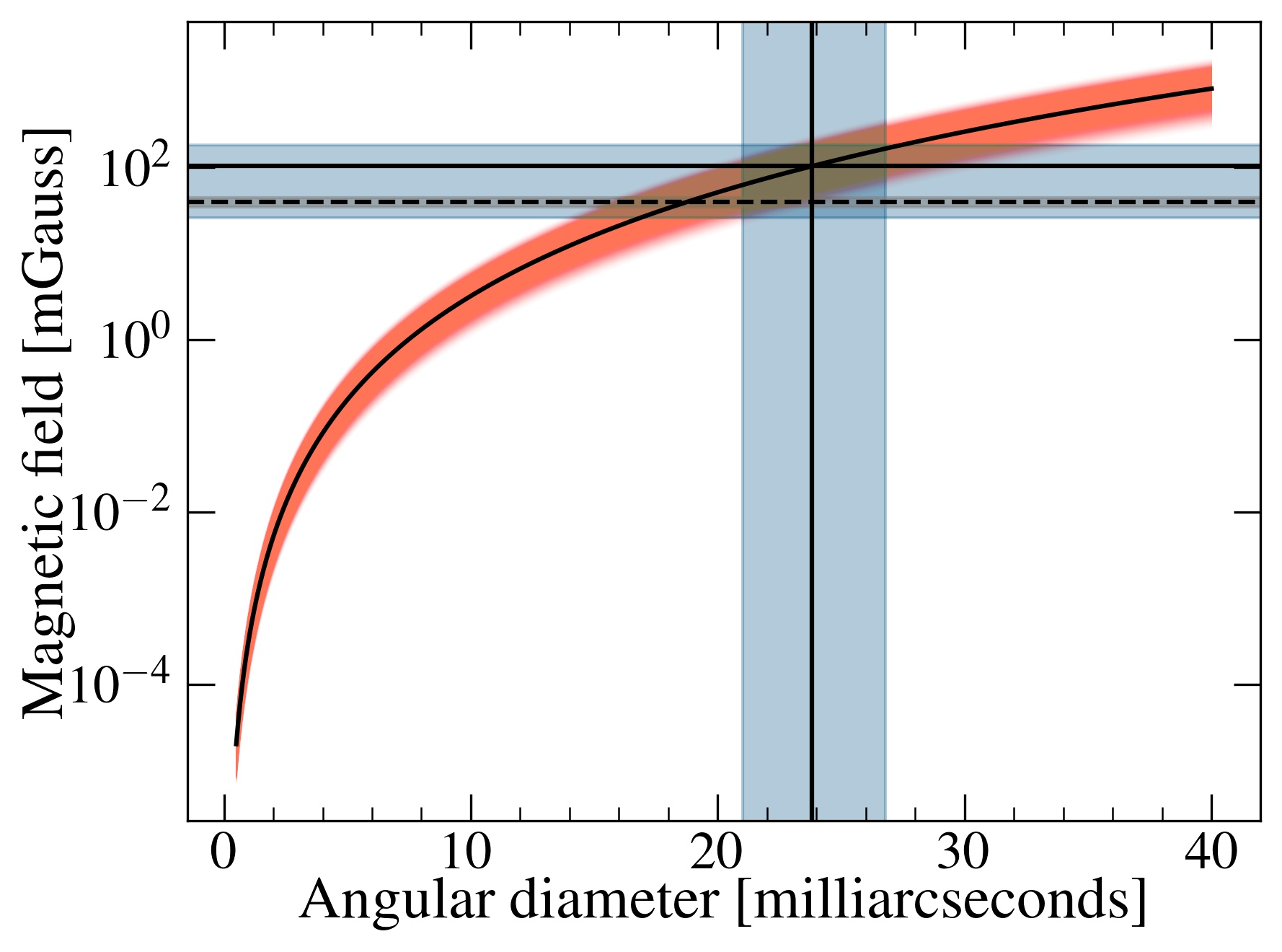}\vspace{0.5em}
    \caption{Variation of the magnetic field strength ($B_{\rm s}$) with angular size ($\theta_{\rm s}$) of the jet knot, according to Equation~(\ref{eq:B}). The red shaded region around the main curve highlights the $1\sigma$ uncertainties on the self-absorption turnover frequency $\nu_{\rm p}$, the corresponding radio flux density $S_0$, and our calculated range for the Doppler factor $\delta_{\rm bp}$. The vertical line at 23.8\,mas indicates the estimated source size on 21 September 2017 (derived from our LBA observation on 23--24 September 2017, assuming constant expansion speed).  The solid horizontal line shows the magnetic field strength ($104^{+80}_{-78}$\,mG) corresponding to the inferred source size. The shaded regions across all the horizontal and vertical lines indicate the $1\sigma$ uncertainties. The dashed black horizontal line at 40\,mG corresponds to the minimum energy field strength $B_{\rm eq}$. Our SSA modelling and LBA size constraint suggest that the jet knot is close to equipartition.
    }
    \label{fig:fig7}
\end{figure}

From the proper motion of the approaching jet knot, \citet{Russell2019} constrained the product $\beta\cos{i}\geq0.49$, implying that the jet speed $\beta\geq0.69$, and $i\leq45^{\circ}$. Using the aforementioned constraints, we defined a uniform distribution of $0.69\leq \beta \leq1.0$ and $1/\sqrt{2}\leq \cos{i} \leq1$, which corresponds to a distribution of {\it i} in the range $0^{\circ}\leq i \leq45^{\circ}$. We calculated the probability density function for $\delta_{\rm bp}$, finding that the 5--95\% likelihood range for $\delta_{\rm bp}$ is 1.0--3.4.  We used our fitted self-absorption turnover frequency and the radio flux density at that frequency (from Section~\ref{sec:Radiospec_SSA}) to determine the magnetic field strength as a function of source size, as shown in Fig.~\ref{fig:fig7}. In Section \ref{sec:Radiospec_SSA}, we calculated the source size on 21 September to be $23.8^{+3.0}_{-2.8}$ mas, which at 4.1\,kpc corresponds to a jet expansion speed, $\beta _{m}$ (in units of $c$), of $0.09\pm0.04$. From equation~(\ref{eq:B}), this implies a magnetic field strength of $104^{+80}_{-78}$\,mG,\footnote{The uncertainty on the magnetic field strength ($B_{\rm s}$) is dominated by the error on the ejection date, which dominates the uncertainty on the source size on 21 September 2017. To reduce the uncertainty on the ejection time in future outbursts, we would need to perform high-cadence VLBI monitoring of the source near the peak of the outburst.} which is consistent with the limits of 10--500\,mG derived from equipartition arguments by \citet{Russell2019}. Our derived magnetic field strength for MAXI J1535--571 is roughly consistent with the values reported for other X-ray binary jets (10\,mG in SS\,433; \citealt{Seaquist1982}, 250\,mG in V404 Cygni; \citealt{Chandra2017}).

The minimum energy magnetic field strength ($B_{\rm eq}$ $\approx40\pm5$ mG) estimated in Section~\ref{sec:Radiospec_SSA} via the formalism of \citet{Fender2006} is consistent (within uncertainties) with the magnetic field strength ($B_{\rm s}$) determined from the LBA size measurement and synchrotron self-absorption theory. This suggests that the transient jet in MAXI J1535--571 is likely to be close to equipartition.

\subsection{Jet opening angle}
\label{sec:opening_angle}
\citet{Russell2019} analysed and fit the proper motion of the discrete jet knot with three different models; ballistic motion, constant deceleration, and ballistic motion plus late-time deceleration. They found that the proper motion of the transient ejecta could be best described by ballistic motion for the first $\sim 260$ days, followed by late-time deceleration.  In this model, the ejection event occurred on MJD $58010.8^{+2.7}_{-2.5}$. The opening angle $\phi_{\rm op}$ can be calculated as \citep{Miller-Jones2006}
\begin{equation}
 \tan\phi_{\rm op} \approx \frac{\theta_{\rm s}}{\mu_{\rm app} (t_{\rm obs}-t_{\rm ej})},
\end{equation}
where $\theta _{s}$ is the size of the jet knot, $\mu_{\rm app}$ is the proper motion of the approaching jet knot, and $(t_{\rm obs}-t_{\rm ej})$ is the time between the ejection event and the observation.

With the measured proper motion of the jet component, $\mu_{\rm app}=47.2\pm1.5$ mas\,day$^{-1}$ \citep{Russell2019}, our LBA size measurement ($\theta_{\rm s}=34\pm1$\,mas) implies an opening angle of $\phi_{\rm op}=4.5\pm1.2\degree$ (independent of the inclination angle of the jet axis). The opening angle is consistent with the upper limit of $\leqslant10\degree$ determined by \citet{Russell2019}. 

Constraints on the jet opening angle $\phi_{\rm op}$ have only been determined for a small sample of BH XRBs \citep{Miller-Jones2006, Rushton2017, Tetarenko2018}, with all except three of these measurements being upper limits. The upper limits of \citet{Miller-Jones2006} range from $<2\degree$ (for the steady hard-state jets in Cygnus X--1) to $\leqslant25.1\degree$ (for V4641 Sgr). The measured values are for Cygnus X-3 ($5\pm0.5^{\circ}$; \citealt{Miller-Jones2006}), XTE J1908+094 ($\sim58^{\circ}$; \citealt{Rushton2017}) and V404 Cygni ($4^{\circ} - 10^{\circ}$; \citealt{Tetarenko2017}). Thus, our opening angle measurement is also in agreement with the typical constraints available for other BH XRBs.

\subsubsection{Lorentz factor}
\begin{figure}
    \includegraphics[width=\columnwidth]{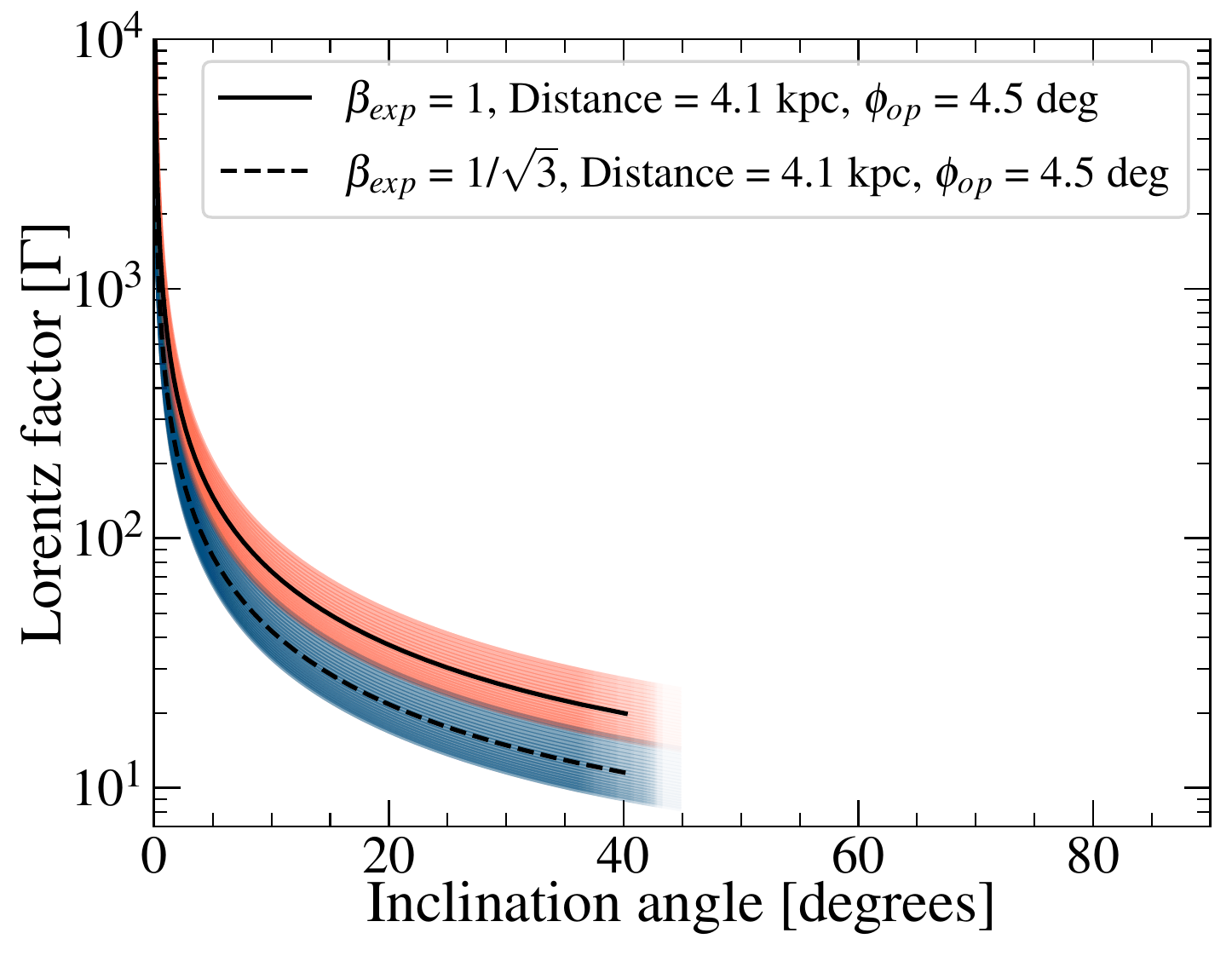}
    \caption{The variation of the bulk Lorentz factor with the inclination angle. The solid and the dashed black lines highlight the expected curves for expansion speeds of c and $c/\sqrt{3}$, respectively. The orange (expansion speed c) and the blue (expansion speed $c/\sqrt{3}$) shaded regions show the effect of incorporating the $1\sigma$ uncertainty on the opening angle and the distance to the source ($4.1^{+0.6}_{-0.5}$ kpc; \citealt{Chauhan2019b}). If the jet is freely expanding, we would predict a bulk Lorentz factor $\Gamma>10$.
    }
    \label{fig:fig6}
\end{figure}

If the jet is not confined, we can use our measurement of the opening angle to constrain the bulk jet Lorentz factor via the formalism given by \citet{Miller-Jones2006} as 

\begin{equation}
\label{eq:gamma}
  \Gamma = \left[1 + \frac{ \beta^{2}_{\rm exp}}{\tan^{2}{\phi_{\rm op}} \sin^{2}{i}}\right]^{1/2},
\end{equation}
where $\beta_{\rm exp}$ is the expansion velocity of the plasma cloud and $i$ is the inclination angle. We assume two different scenarios for the expansion of the jet knot. In one case, we assume the knot is expanding at the relativistic sound speed, $c/\sqrt{3}$. Alternatively, we assume the plasma knot is expanding freely with speed $c$ \citep{Miller-Jones2006}. We determine $\Gamma$ for all permissible values of $i$ ($\leq45^{\circ}$; \citealt{Russell2019}) as shown in Fig.~\ref{fig:fig6}. For all allowed values of $i$, $\Gamma$ $>10$ (Fig.~\ref{fig:fig6}).

The commonly-assumed range for the Lorentz factor in transient XRB jets is $2<\Gamma<5$ \citep{Miller-Jones2006}; significantly lower than that inferred for MAXI J1535--571 assuming free expansion. Although some recent studies \citep[e.g.,][]{Casella2010, Tetarenko2019, Tetarenko2021} have suggested that the compact jet Lorentz factors may be significantly higher, transient jet Lorentz factors are typically poorly constrained. \citet{Fender2003} demonstrated that if the distance to the source is not accurately known, then for any significantly relativistic jet, we can only determine a lower limit on $\Gamma$ from the proper motions of the bipolar transient ejecta.

From the constraints on $\beta\cos{i}$ derived for the transient jets of MAXI J1535--571 by \citet{Russell2019}, high Lorentz factors would imply a very low inclination angle, which is relatively improbable, and inconsistent with the disk inclination angles determined from fitting the X-ray reflection spectrum \citep[which themselves differ significantly;][]{Miller2018,Xu2018}. However, discrepancies have been observed between the jet inclination angle, and the disk inclination angle estimated using reflection modelling \citep[e.g.,][]{Atri2020, Xu2020}. These discrepancies could be due to systematic effects not accounted for in the reflection modelling, or to intrinsic misalignments in the system geometry \citep[e.g.][]{Maccarone2002}.  Nonetheless, the inferred dependence of Lorentz factor on inclination angle shown in Figure~\ref{fig:fig6} disagrees with that derived from the observed jet proper motions \citep{Russell2019}, so we do not favour such a low inclination angle for the jets.

Given the above, we suggest that the jet knot in MAXI J1535--571 is not likely to be freely expanding, but is instead externally confined. This is in agreement with the relatively low expansion speed of $\beta_{\rm m}=0.09\pm0.04$ derived from the measured source size in Section~\ref{sec:magnetic_field}. Our calculated $\beta _{\rm m}$ is consistent with the limit ($<0.18$\,c) determined by \citet{Russell2019} for the expansion speed. Additionally, \citet{Tetarenko2017} estimated similarly low expansion speeds of 0.01-- 0.1\,$c$ for the transient jet knots in V404 Cygni, and found that the jet knots were externally confined. Potential confinement mechanisms could include the disc magnetic field, the thermal gas pressure from the surrounding medium, or inertial confinement by an outflowing wind \citep[see, e.g.,][]{Begelman1984, Ferrari1998, Miller-Jones2006}. With the available results, we cannot discriminate between the above-mentioned confinement mechanisms. Confinement due to the thermal gas pressure from the surrounding medium is certainly plausible, because the jet knot would have travelled a significant distance from the black hole between the time of ejection and the time of our observations.

\subsection{Future studies with the SKA-Low}
Our study showcases the potential contributions that SKA-Low (frequency range 50--350\,MHz) could make to the study of radio jets from black hole X-ray binaries. MAXI J1535--571 was relatively bright, and hence accessible to the current SKA precursor facilities.  Most BH XRB jets are somewhat fainter [e.g., EXO 1846--031 \citep{Williams2019}, MAXI J1820+070 \citep{Bright2020}, and MAXI J1803--298 \citep{Espinasse2021}], peaking at a few tens of mJy at GHz frequencies. Such faint systems would require (sub-)mJy-level sensitivity for us to detect them at low radio frequencies, and could not be effectively probed by MWA.  While lower-power transient jets may initially be smaller, and hence would evolve more rapidly, their spectra should nonetheless evolve to lower frequencies with time \citep{Laan1966}, albeit with fainter peak flux densities. Provided they are observed sufficiently early in their evolution, the higher expected sensitivity of SKA-Low (14--26\,$\mu$Jy\,beam$^{-1}$\,hr$^{-1/2}$\footnote{\url{https://www.astron.nl/telescopes/square-kilometre-array/}}) should therefore allow us to study the transient jets from faint BH XRBs, as well as neutron star XRBs, whose faintness has to date precluded the kinds of detailed studies performed on black holes. The SKA-Low will therefore enable the detailed exploration of accretion-ejection coupling across a broader range of stellar-mass compact objects \citep[see, e.g.,][for further details]{Corbel2015}.
  
\section{CONCLUSIONS}
\label{sec:conclusions}
In this work, we have conducted a multi-wavelength study of the transient jet from the black hole candidate XRB MAXI J1535--571. We presented new results from MWA, UTMOST and LBA, and included previously-published results from ASKAP and ATCA, collectively providing spectral coverage from 0.12--19\,GHz. During our campaign, we made the first MWA detection of a transient radio jet from a black hole XRB, detecting MAXI J1535--571 at a significance $>10 \sigma$. Using our LBA observation on 23 September 2017, we constrained the source size to $34\pm1$ mas, which corresponds to a physical size of $139^{+21}_{-17}$\,AU, calculated using a source distance of $4.1^{+0.6}_{-0.5}$\,kpc from \citet{Chauhan2019b}. The size constraints allowed us to calculate the jet opening angle to be $4.5\pm1.2\degree$. Given the large bulk Lorentz factor that would be implied in the case of a freely-expanding jet, we infer that the jet knot is likely to be externally confined.

Our broad-band spectrum on 21 September 2017 indicates the presence of a low-frequency spectral turnover, whose peak frequency is strongly constrained by the MWA observations. The detected low-frequency turnover is likely due to synchrotron self-absorption. We fitted the broadband spectrum with a self-absorption model, and calculated the power-law index of the energy distribution for the relativistic electrons of the source ($\beta = 1.90\pm0.04$), the turnover frequency ($\nu _{p} = 0.32\pm0.01$ GHz), and the corresponding peak flux density ($S_{0} = 882\pm56$ mJy).

We further used our LBA size constraint along with the turnover frequency and the peak flux density obtained from the SSA model fitting to calculate the magnetic field strength ($104^{+80}_{-78}$ mG) and the minimum energy ($6.5\pm2.5\times10^{41}$ erg) of the jet knot. These estimates are consistent with the values estimated for MAXI J1535--571 by \citet{Russell2019}, and with canonical XRB values.  

Finally, our study highlights the combined capabilities of the Australian suite of radio telescopes including MWA, UTMOST, ASKAP, ATCA and LBA, which can provide sensitive and simultaneous broadband coverage of radio jets in XRBs. This will be significantly augmented over the coming years as we move into the era of the SKA.

\section*{Acknowledgements}
We thank the referee for their valuable comments. This scientific work makes use of the Murchison Radio-astronomy Observatory, operated by CSIRO. We acknowledge the Wajarri Yamatji people as the traditional owners of the Observatory site. Support for the operation of the MWA is provided by the Australian Government (NCRIS), under a contract to Curtin University administered by Astronomy Australia Limited. The Australia Telescope Compact Array and Long Baseline Array are both part of the Australia Telescope National Facility, which is funded by the Australian Government for operation as a National Facility managed by CSIRO. We acknowledge the Pawsey Supercomputing Centre which is supported by the Western Australian and Australian Governments. JCAM-J is the recipient of Australian Research Council Future Fellowship (project number FT140101082) and GEA is the recipient of an Australian Research Council Discovery Early Career Researcher Award (project number DE180100346) funded by the Australian Government. NHW is the recipient of Australian Research Council Future Fellowship (project number FT190100231). DLK was supported by NSF grant AST--1816492. TDR acknowledge financial contribution from ASI-INAF n.2017-14-H.0, an INAF main stream grant. SWD acknowledges an Australian Government Research Training Program scholarship administered through Curtin University. DA acknowledges support from the Royal Society. SC and DLK were supported by NSF grant AST-1412421. GRS acknowledges support from Natural Sciences and Engineering Research Council of Canada (NSERC) Discovery Grant (RGPIN-06569-2016).

\bibliographystyle{pasa-mnras}
\bibliography{References}

\end{document}